# Agency and legibility for artists through Experiential AI

Drew Hemment[†]
Edinburgh College of Art
University of Edinburgh
Edinburgh
drew.hemment@ed.ac.uk

Matjaz Vidmar
Engineering
University of Edinburgh
Edinburgh
matjaz.vidmar@ed.ac.uk

Daga Panas
Informatics
University of Edinburgh
Edinburgh
daga.panas@ed.ac.uk

Dave Murray-Rust
Industrial Design Engineering
TU Delft
Delft
d.s.murray-rust@tudelft.nl

Vaishak Belle
Informatics
University of Edinburgh
Edinburgh
vbelle@ed.ac.uk

Ruth Aylett
Computer Science
Heriot-Watt University
Edinburgh
r.s.aylett@hw.ac.uk

## ABSTRACT

Experiential AI is an emerging research field that addresses the challenge of making AI tangible and explicit, both to fuel cultural experiences for audiences, and to make AI systems more accessible to human understanding. The central theme is how artists, scientists and other interdisciplinary actors can come together to understand and communicate the functionality of AI, ML and intelligent robots, their limitations, and consequences, through informative and compelling experiences. It provides an approach and methodology for the arts and tangible experiences to mediate between impenetrable computer code and human understanding, making not just AI systems but also their values and implications more transparent, and therefore accountable. In this paper, we report on an empirical case study of an experiential AI system designed for creative data exploration of a user-defined dimension, to enable creators to gain more creative control over the AI process. We discuss how experiential AI can increase legibility and agency for artists, and how the arts can provide creative strategies and methods which can add to the toolbox for human-centred XAI.

## KEYWORDS

## 1  Introduction

Recent developments in diffusion models and large language models have powered a new generation of AI tools, opening unprecedented opportunities for artistic creation, with profound implications for the arts and society. However, this latest generation of tools, while powerful, perform narrow and prescribed tasks, where the user remains oblivious as to the mechanics of creation and retains little control of it.

Experiential AI [1] studies and develops accessible and legible tools for the arts, and creative methods for explainable AI (XAI). Work in this field develops methods and tools to make data-driven AI and machine learning tangible, interpretable, and accessible to the intervention of a user or audience. Such an approach can use tangible experiences to communicate black-boxed decisions and nuanced social implications in engaging, experiential ways, with high fidelity to the concepts. We propose this practice offers new modalities of explanation, that open up algorithms, the science behind them, and their potential impacts in the world to scrutiny and debate [2]. It foregrounds embodied experience and situated meaning in a similar way to soma design [3], and is distinguished from other movements in human-centred and creative AI by a focus on tangible explanation as a means to support public literacies.

Explainable AI (XAI) is the leading current approach within the AI community investigating how the decisions and actions of machines can be made explicable to human users [4]. We argue that despite significant achievements in XAI, the field is limited by a focus on technical explanations. Turning to the arts for inspiration, we propose that artistic methods can enhance the legibility of intelligent systems and scaffold understanding of AI. Artistic experiences can enable a holistic engagement, that is particularly relevant to understand the nuanced social implications of deployed AI technology, alongside and not isolated from socio-technical applications. The arts can engage people emotionally, cognitively and tangibly with the large-scale effects of pervasive AI deployments. And yet, the arts are not instruction, and it would be wrong to instrumentalise the arts, either for science communication or system design. Therefore, we develop experiential AI as a cross- or trans-disciplinary methodology, that is complementary and additional to the practice of artists, and that has a specific focus on orchestrating and brokering between disciplines.

Our research [2] has identified the following six gaps or areas for development in XAI that experiential AI can help address:

1. Providing explanations from human points of view
2. Looking beyond model explanations to address the entirety of the AI system
3. Connecting technical aspects to higher-level dimensions of AI
4. Accounting for a wider range of stakeholders for systems deployed in social situations
5. Engaging with the imaginaries surrounding AI
6. Deeper engagement in material and ideological realities

## 2   Methods

The New Real group [5] conducts research that brings professional artists together with scientists, engineers and applied ethicists to create near future scenarios in which machine learning systems, social robots or other technologies can be designed, deployed and tested as experiences, in the form of interfaces, installations, performances, situations, interactions. An experiential AI system was developed for and with artists, as a part of a collaborative journey in creating and presenting a number of artworks, each one addressing issues in AI such as authorship, misinformation, harmful bias, or energy use. An open prototyping methodology [6] is used to broker interactions between artists, ethicists, data scientists, design researchers, audiences, data, AI algorithms, and other mediating technologies. The platform's functionality and the configuration of algorithms, as well as their structuring into 'tools' was articulated and refined through a series of developmental workshops negotiating between participants' requirements and available technical solutions. System design was also informed by an assessment of currently available creative and generative AI tools, which highlighted a gap for accessible tools that are configurable and offer granular control. The resulting platform (Figure 1) was developed to encompass a variety of machine learning and other computational techniques to enable human-machine exploration and experimentation with the data at hand. The platform generates visual, textual and numerical outputs that the artists can use as material for art pieces, and, by extension, experiential explanations. The toolset was configured for probing the model, to expose the AI, and for greater accessibility, legibility and agency. The artists built on this data exploration through the development of configurations of interfaces, artefacts, narratives and interactions which comprise the 'artworks' that were deployed to engage and delight online and in person audiences (Figure 2). The New Real team used curatorial experience and know-how, and engaged an exhibition designer, to frame the works for display and produce an exhibition, presented as part of Ars Electronica Festival 2022 (Figure 3).

## 3   Results

A key empirical result was a set of tools for artists to create models based on their own data and, to construct and explore customised dimensions-of-interest in their model's latent space, instead of relying solely on already trained models pre-configured to serve outputs in a manner that provides

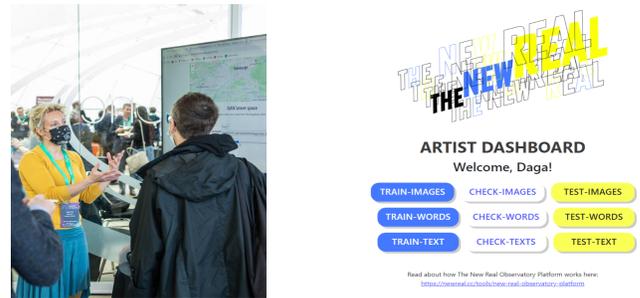
Figure 1. *The New Real Observatory* experiential AI platform by The New Real 2022 and demo at Scottish AI Summit 2022.

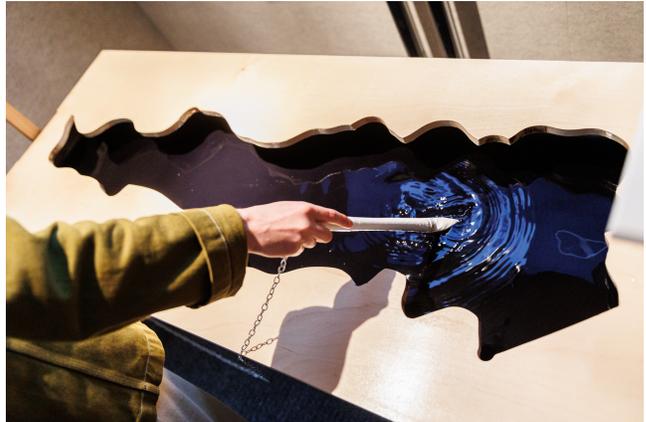
Figure 2. *Photographic Cues* by Keziah MacNeill 2022 at The New Real Pavilion, Ars Electronica, Linz 2022. Photo by DieFotoFrau

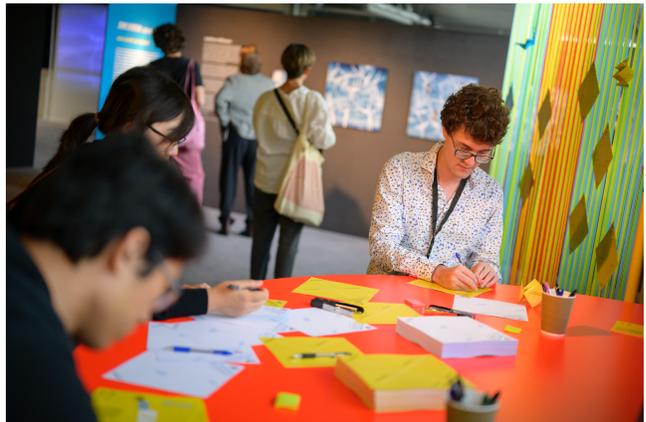
Figure 3. *The New Real Pavilion and Research Hub* by The New Real 2022 at Ars Electronica, Linz 2022. Photo by DieFotoFrau.

minimal freedom to the artist. To enable users to define a 'dimension', two pipelines were developed around the training or fine tuning of a model and subsequent processing of additional information: an 'images' pipeline centered around a generative model; and a 'words' pipeline centered around a word embedding model. A novel 'SLIDER' – Shaping Latent-spaces for Interactive Dimensional Exploration and Rendering – tool was developed as the final stage within each of the pipelines, allowing the artist to probe their model for an output at a specific position in the newly-constructed dimension.

Each pipeline has a distinct core creative medium, as well as different model architectures and algorithm affordances, and the implementation of the human-machine interaction was accordingly different in each. In the visual processing pipeline, the user submits two separate sets of images, forming distinct, internally consistent classes, in order to train and map a dimension between them, and is then able to generate new images upon probing. In the semantic processing pipeline, the user needs only a single corpus of text for training, they can then construct multiple 'sliders' for the model by entering a list of words, and are then able to generate associations by entering a base word for probing, rather than words being produced ad novo. In addition, the semantic pipeline is equipped with a visualisation tool, presenting the latent space as a flattened, annotated 'point cloud'.

The artists reported that the framing of dimensions with their own data, and then extracting model outputs via the slider, provided them with granular control over the output of the model, and this enabled them to better explore and experience how different inputs shape the outputs generated by the system. One artist reported the combination of the visualisation tool with word to vector transformations provided:

> "a really unique output that falls between language modelling and text to image modelling … offering an exciting view into the process and word associations that occur in the high dimensional latent space of machine learning."

Using the exploration tools the artists were able to probe the model, expose semantic associations and biases, and explore the AI's 'understanding' of the submitted dimensional dataset. They used the platform to find fresh perspectives on their dimension-of-interest that they could build on in the development of an artistic work:

> **Artist 1** exposed the link between computation and perception.
> **Artist 2** reframed and disrupted value systems in machine vision.
> **Artist 3** explored future scenarios combining statistical and qualitative methods.
> **Artist 4** asked profound questions of AI, through a speculative, material future.

The artworks offer novel perspectives on the AI, data, and the mutual shaping of technology and society. They include imagery, narrative, interaction, sound and other media which interpret the output of the model or otherwise bridge the statistical, mechanised AI and organic, intuitive human intelligence. They produce situated, embodied and intuitive meaning around algorithms and the effects of their deployments. This can help to increase understanding of what the AI is, and how various things are being characterised by/as data. Each artwork is understood as an instance of the Experiential AI system, and offers a distinct frame of reference for characterising that system, illuminating different modalities of interaction between the artist and algorithm, of tangible experience for audiences, and of human and machine learning.

## 4  Discussion

We have seen how experiential AI can generate cultural experiences for audiences, and also make AI systems accessible to human understanding. We saw how this offered the practitioner creative control and agency in co-creative experiments with AI, going beyond the current paradigm of prompt engineering. We also say how this can make the opaque mechanisms and meanings of AI artefacts and algorithms transparent to those who interact with them. Beyond communicating current knowledge, an experiential approach can generate new understanding on AI systems – their operations, limitations, peculiarities and implications.

We envision future research to develop experiences with explanatory skill for various aspects of the life cycle of AI systems, from data collection, systems design, algorithm selection and deployment, through to the interests and ideologies vested in their decisions and the social implications that follow. This includes not just the models at their core, but also the data collection and processing that gives rise to them, the way the system has been commissioned and designed, how automated decision making is situated, what the model actually does, as well as the relations between the system and the subjects of its decisions.

Such approaches are needed to assist designers in approaching how AI is situated in the world, to increase the range of people engaged in shaping the field, and to restore the basis for accountability. Holistic questions can then be asked of the entanglements of humans and machines, going beyond model interpretability, such as how does AI challenge our world view, how does it shape human experience and relationships, and how we can avoid anthropomorphism and misplaced trust in AI. Such experiential interventions can work to reach new audiences, to increase the agency of people impacted by these systems, and to create spaces for debate and engagement with populations outside the technical centre.

Our work leads us to the following recommendations:
1. Connect the technology to social goals;
2. Combine the efforts of artists and XAI practitioners, while preserving an autonomy for the practice of individual contributors;
3. Ground the art in current science;
4. Develop creative expressions with high fidelity to the technical concepts;
5. Provide an interface to the public and policy debates;
6. Enable rigorous evaluation.

## ACKNOWLEDGMENTS
Thanks to Inés Cámara Leret, Keziah MacNeill, Adam Harvey, Alex Fefegha, Roy Luxford, Ricarda Bross, Andy McGregor, Evan Morgan, Mario Antonioletti and Miriam Walsh. The New Real at University of Edinburgh is a partnership with The Alan Turing Institute and Edinburgh's Festivals. Funded by Engineering & Physical Sciences Research Council/Turing 2.0, Arts & Humanities Research Council and Creative Scotland.